\documentclass[prd,floats,floatfix,nofootinbib]{revtex4}
\usepackage[dvips]{graphicx}
\usepackage{graphics}
\usepackage{amssymb}
\usepackage{amsmath}
\usepackage{color}

\begin{document}

\title{The Vacuum Displacement Principle:\\
Theoretical Framework and Local Phenomenology}

\author{Rodrigo Maier\footnote{rodrigo.maier@uerj.br}} 

\affiliation{
Departamento de F\'isica Te\'orica, Instituto de F\'isica, Universidade do Estado do Rio de Janeiro,\\
Rua S\~ao Francisco Xavier 524, Maracan\~a,\\
CEP20550-900, Rio de Janeiro, Brasil
}

\date{\today}

\begin{abstract}
We present a modified gravitational framework in which the standard Einstein field equations are sourced by a classical matter sector coupled to a Higgs-type scalar field $\chi$ modeling a dynamic vacuum substrate. By introducing a phenomenological covariant coupling, we implement a physical displacement principle where massive baryonic matter drives the vacuum field away from its vacuum expectation value. We show that this coupling leads to a field-dependent modulation of a particle's inertial rest mass alongside a spatial fifth force, yielding localized violations of the Einstein Equivalence Principle while preserving universal free fall for fundamental point masses. In the weak-field, non-relativistic limit, this interaction manifests as a Yukawa-type correction to the Newtonian potential. We test the viability of this framework against local gravitational constraints, demonstrating that precision Eötvös parameter limits provide the primary operational bound, whereas long-range scalar interactions leave secular planetary perihelion precession unaffected. Finally, we model the steady-state, non-relativistic spherical accretion of dust, demonstrating how the interplay between fifth-force acceleration and mass modulation uniquely reshapes the infalling fluid's density and velocity profiles.
\end{abstract}

\maketitle

\section{Introduction}

The cornerstone of modern gravitation, General Relativity, is built upon the Weak Equivalence Principle (WEP), which posits the universality of free fall and the fundamental identity of inertial and gravitational mass. While this framework has passed every experimental test within the Solar System and beyond\cite{Will:2014kxa}, the emergence of the standard cosmological model has introduced profound conceptual and observational challenges. The staggering discrepancy between the observed cosmological constant and the theoretical zero-point energy of quantum fields\cite{Weinberg:1988cp} strongly indicates that the coupling between spacetime geometry and the quantum vacuum is fundamentally misunderstood. While modern field theory describes the vacuum as a dynamical substrate characterized by spontaneous symmetry breaking and phase transitions\cite{Weinberg:1996kr}, General Relativity treats it as a passive geometric constant. This conceptual disparity raises a fundamental question: if the vacuum is a dynamical entity, must it not respond locally to the presence of matter, and if so, how does this interaction redefine the laws of motion?

A compelling microphysical motivation for a dynamical vacuum arises from the framework of Running Vacuum Models. Recent results in Quantum Field Theory in curved spacetime demonstrate that the vacuum energy density is not a rigid constant but a running quantity that evolves with the characteristic energy scale of the universe, typically associated with the Hubble rate\cite{SolaPeracaula:2022hpd,SolaPeracaula:2023swx}. These models show that the running of the vacuum is a formal consequence of the renormalization group evolution of the effective action, where the vacuum and matter sectors are fundamentally coupled through the gravitational field\cite{Moreno-Pulido:2020anb}. This suggests that the vacuum is a responsive substrate capable of exchanging energy-momentum with the matter sector, providing a phenomenological justification for the responsiveness of the vacuum to its local and global environment.

From a phenomenological perspective, an interacting vacuum-matter sector offers a unified resolution to several persistent cosmological anomalies. The Coincidence Problem -- the question of why the energy densities of dark matter and dark energy are of the same order today -- is naturally addressed if these sectors are coupled, allowing one to track the other\cite{Zimdahl:2001ar,Wang:2016lxa}. Furthermore, such interactions provide new degrees of freedom to alleviate the $H_0$ tension, as a dynamical vacuum can shift the expansion rate in a way that reconciles early and late-universe measurements\cite{DiValentino:2021izs}.

In this paper, we propose a novel framework based on the \textit{Vacuum Displacement Principle}. Unlike standard models where matter exists within a passive vacuum backdrop, we posit that baryonic matter displaces the vacuum substrate. Formally, this implies that matter acts as a localized impurity that pushes the vacuum field away from its equilibrium state. To model this vacuum substrate from a phenomenological point of view, we employ an ultra-light dark scalar field $\chi$ characterized by a spontaneous symmetry breaking potential. However, we stress however that $\chi$ is not the Standard Model electroweak Higgs field, nor does it belong to the electroweak particle sector\cite{Higgs:1964pj,Englert:1964et,Guralnik:1964eu}. Instead, it represents a dark energy/quintessence-like scalar sector coupling locally to matter sources. The choice of a symmetry breaking potential is motivated by its ability to define a clear vacuum stiffness via the self-coupling $\lambda$ and a stable ground state via its Vacuum Expectation Value (VEV) $v$.

This work is organized as follows: In Section~\ref{sec:theory}, we define the theoretical framework and establish the modified Klein-Gordon field equations for the scalar vacuum substrate. In Section~\ref{sec:eom_wep}, we derive the modified equations of motion for test particles and analyze the mechanism driving Equivalence Principle violations. Section~\ref{sec:newtonian_limit} explores the coupled Newtonian limit, demonstrating the emergence of a Yukawa-type effective potential and evaluating WEP bounds. In Section~\ref{sec:vacuum_solutions}, we examine vacuum solutions under spherical symmetry, recovering the Schwarzschild metric in empty space and constraining the coupling via planetary perihelion precession. In Section~\ref{sec:steady_state_accretion}, we model steady-state, non-relativistic fluid accretion onto compact objects, showing how the vacuum displacement which follows from our model alters fluid density and velocity profiles. 
Finally, Section~\ref{sec:discussion} offers our concluding remarks and discusses future avenues.

\section{Theoretical Framework}
\label{sec:theory}

We formulate the theoretical framework in a four-dimensional spacetime with metric signature $(-,+,+,+)$. The gravitational field equations take the standard form:
\begin{equation}
G_{\mu\nu} = \kappa^2 \left( T_{\mu\nu} + V_{\mu\nu} \right),
\end{equation}
where $\kappa^2 \equiv 8\pi G = M_{\text{Pl}}^{-2}$, $T_{\mu\nu}$ represents the stress-energy tensor of the matter sector, and $V_{\mu\nu}$ denotes the vacuum stress-energy tensor associated with a real scalar field $\chi$. The action governing the scalar vacuum substrate is given by
\begin{equation}
\label{eq:action_chi}
\mathcal{S}_{\chi} = -\int d^4x \sqrt{-g} \left[ \frac{1}{2}g^{\mu\nu}\nabla_\mu \chi \nabla_\nu \chi + U(\chi) \right],
\end{equation}
where the scalar potential adopts a Higgs-type spontaneous symmetry breaking form:
\begin{equation}
\label{eq:potential}
U(\chi) = \frac{1}{4}\lambda\left(\chi^2 - v^2\right)^2.
\end{equation}
In the above $v$ denotes the VEV in the absence of matter, and $\lambda$ is a dimensionless self-coupling parameter controlling the stiffness of the vacuum potential. Varying Eq.~\eqref{eq:action_chi} with respect to the metric yields the canonical scalar energy-momentum tensor
\begin{equation}
V_{\mu\nu} \equiv -\frac{2}{\sqrt{-g}}\frac{\delta \mathcal{S}_\chi}{\delta g^{\mu\nu}} = \nabla_\mu \chi \nabla_\nu \chi - g_{\mu\nu} \left[ \frac{1}{2} \nabla^\alpha \chi \nabla_\alpha \chi + U(\chi) \right].
\end{equation}

The matter-vacuum coupling is now formally implemented by allowing a non-trivial energy-momentum exchange vector $Q^\nu$ between the matter and vacuum sectors:
\begin{align}
\nabla_\mu T^{\mu\nu} &= Q^\nu, \label{eq:matter_cons}\\
\nabla_\mu V^{\mu\nu} &= -Q^\nu, \label{eq:vac_cons}
\end{align}
such that the total energy-momentum tensor is conserved, namely, $\nabla_\mu (T^{\mu\nu} + V^{\mu\nu}) = 0$, guaranteeing compatibility with the contracted Bianchi identities $\nabla_\mu G^{\mu\nu} = 0$.

To model the physical scenario where matter acts locally to displace the vacuum, we adhere to a phenomenological covariant coupling proportional to the trace of the matter stress-energy tensor, that is,
\begin{equation}
\label{eq:coupling_Q}
Q^\nu = \frac{\beta}{M_{\text{Pl}}} T \nabla^\nu \chi,
\end{equation}
where $\beta$ is a dimensionless coupling strength parameter. By tying the coupling directly to $T$, classical electromagnetic fields remain decoupled from the scalar substrate, ensuring that photon propagation conforms to standard null geodesics. Although quantum loop corrections\cite{Adler:1976zt,Collins:1976yq,Crewther:1972kn} generate a non-zero trace via the conformal anomaly $\langle T^\mu_\mu \rangle \neq 0$, this loop-suppressed interaction is supposed to be negligible in the
energy scales analyzed in this first analysis.

The physical picture behind this interaction is intuitive: baryonic matter acts as a localized impurity embedded within the scalar vacuum substrate. The potential $U(\chi)$ defines an equilibrium state at $\chi = \pm v$. The presence of matter exerts a localized displacement pressure that pushes the scalar field away from its VEV toward $\chi = 0$, forcing the vacuum into a higher potential energy state. 

The field dynamics governing the vacuum substrate follow directly from the non-conservation equation \eqref{eq:vac_cons} together with \eqref{eq:coupling_Q}. Evaluating the covariant divergence of $V^{\mu\nu}$ the modified Klein-Gordon equation reads
\begin{equation}
\label{eq:modified_kg}
\Box \chi - \lambda \chi \left(\chi^2 - v^2\right) + \frac{\beta}{M_{\text{Pl}}} T = 0.
\end{equation}
For ordinary non-relativistic matter with $T \approx -\rho < 0$, the interaction term $\frac{\beta}{M_{\text{Pl}}} T \approx -\frac{\beta}{M_{\text{Pl}}} \rho$ drives the scalar field away from its VEV providing a concrete realization of the Vacuum Displacement Principle.

\section{Equations of Motion and Equivalence Principle}
\label{sec:eom_wep}

In the present framework, the interaction vector $Q^\nu$ governs the exchange of energy-momentum between the matter sector and the vacuum substrate. To derive the resulting equations of motion, we model a non-relativistic dust fluid ($T^{\mu\nu} = \rho u^\mu u^\nu$ with $T = -\rho$) representing a collection of test particles. The non-conservation equation $\nabla_\mu T^{\mu\nu} = Q^\nu$ implies that matter worldlines deviate from standard metric geodesics. In fact, from \eqref{eq:matter_cons} and \eqref{eq:coupling_Q} it follows
\begin{equation}
\label{eq:conservation}
\nabla_\mu \left(\rho u^\mu u^\nu\right) = -\frac{\beta}{M_{\text{Pl}}} \rho \nabla^\nu \chi.
\end{equation}
Projecting Eq.~\eqref{eq:conservation} parallel and orthogonal to the four-velocity $u_\nu$ reveals two distinct physical phenomena: a field-dependent modulation of rest mass and a fifth-force spatial acceleration.

First, the longitudinal projection is obtained by contracting Eq.~\eqref{eq:conservation} with $u_\nu$. Using $u_\nu u^\nu = -1$ and the kinematic orthogonality condition $u_\nu u^\mu \nabla_\mu u^\nu = 0$ one may show that
\begin{equation}
\label{eq10}
\frac{d\rho}{d\tau} + \rho \nabla_\mu u^\mu = \frac{\beta}{M_{\text{Pl}}} \rho \frac{d\chi}{d\tau},
\end{equation}
where $d\rho/d\tau \equiv u^\mu \nabla_\mu\rho$ is the directional derivative along the worldline. For a coherent fluid element or test body with negligible expansion ($\nabla_\mu u^\mu \approx 0$), integrating this relation along the trajectory yields the field-dependent rest-mass relation:
\begin{equation}
\label{eq:mass_chi}
m(\chi) = m_0 \exp\left( \frac{\beta}{M_{\text{Pl}}} \chi \right),
\end{equation}
where $m_0$ is a constant integration parameter representing the bare mass in the symmetric vacuum state $\chi = 0$.

Second, the transverse projection is isolated using the projection tensor $h^{\mu\nu} \equiv g^{\mu\nu} + u^\mu u^\nu$ (which satisfies $u_\mu h^{\mu\nu} = 0$). Contracting Eq.~\eqref{eq:conservation} with $h^\alpha_{~\nu}$ yields the modified geodesic equation
\begin{equation}
\label{eq:eom}
\frac{d^2x^\alpha}{d\tau^2} + \Gamma^\alpha_{\mu\sigma} \frac{dx^\mu}{d\tau} \frac{dx^\sigma}{d\tau} = -\frac{\beta}{M_{\text{Pl}}} \left(\delta^{\alpha}_{~\nu} + u^\alpha u_\nu\right) \partial^\nu \chi.
\end{equation}
The right-hand side of Eq.~\eqref{eq:eom} represents a fifth force mediated by orthogonal gradients of the vacuum substrate field $\chi$.

This $1+3$ spacetime decomposition clarifies the dual role of the interaction: the scalar field gradient parallel to a particle's trajectory drives the dynamical evolution of its rest mass $m(\chi)$, whereas the orthogonal gradient exerts a dynamical fifth force. Consequently, inertial mass becomes an effective, environment-dependent quantity governed by the local state of the vacuum substrate.

Crucially, because the constant bare mass $m_0$ cancels out entirely in the acceleration equation \eqref{eq:eom}, universal free fall is preserved for ideal point-like test particles, satisfying the WEP. However, the position-dependent mass $m(\chi)$ causes local physical scales—such as particle rest masses and bound-state energy levels—to vary across spacetime. This breaks Local Position Invariance (LPI), leading to a direct violation of the broader Einstein Equivalence Principle (EEP)\cite{Will:2014kxa}.

\section{The Coupled Newtonian Limit}
\label{sec:newtonian_limit}

To investigate the weak field behavior, we linearize the system around the vacuum state $\chi = v$ and a Minkowski background $\eta_{\mu\nu}$. We define the field and metric perturbations as $\chi = v + \delta\chi$ and $g_{00} = -(1 + 2\Phi)$, where $\Phi$ is the Newtonian gravitational potential. For a static, nonrelativistic source, the matter contribution is dominated by the rest energy density, such that $T_{00} \approx \rho$ and the trace evaluates to $T \approx -\rho$. 

The vacuum energy contribution $V_{00}$ is derived from its field energy density
\begin{equation}
V_{00} = \frac{1}{2}(\nabla\chi)^2 + U(\chi).
\end{equation}
In the linear approximation, the spatial  gradient term is of second order and can be neglected. Expanding the potential $U(\chi)$ around the VEV via a Taylor series
one may show that
\begin{equation}
\label{eq:ueff}
U(v + \delta\chi) \approx \frac{1}{2}m_\chi^2 \delta\chi^2,
\end{equation}
where $m_\chi^2 \equiv U''(v) = 2\lambda v^2$ defines the effective mass of the scalar field excitations.

From Eq.~\eqref{eq:ueff}, the gravitational contribution of the perturbed vacuum energy density $\delta \rho_{\text{vac}} \sim \mathcal{O}(\delta\chi^2)$ is second-order and therefore negligible compared to the matter density $\rho$ in the Poisson equation. In the modified Klein-Gordon equation \eqref{eq:modified_kg}, however, the linear coupling term acts as a direct source, yielding the coupled system:
\begin{align}
\nabla^2 \Phi &= 4\pi G \rho, \label{eq:poisson_full} \\
(\nabla^2 - m_\chi^2) \delta\chi &=  \frac{\beta}{M_{\text{Pl}}} \rho. \label{eq:kg_full}
\end{align}
Solving Eq.~\eqref{eq:kg_full} for a point mass $M$ at the origin -- so that $\rho = M\delta^3(\mathbf{r})$ -- we obtain the following Yukawa displacement profile
\begin{equation}
\delta\chi(r) = - \frac{\beta M}{4\pi r M_{\text{Pl}}} e^{-m_\chi r}.
\end{equation}
Although the mass parameter $M$ itself depends on $\chi$ via $m(\chi) = m_0 e^{\beta \chi / M_{\text{Pl}}}$, we approximate $M(\chi) \approx M(v) \equiv M$ in the source term; corrections proportional to $M(\chi) \approx M(v)(1 + \frac{\beta}{M_{\text{Pl}}} \delta \chi)$ generate higher-order terms in $\beta/M_{\text{Pl}}$ that are naturally suppressed in this linear weak-field limit.

The effective acceleration of a test particle is dictated by the modified geodesic equation \eqref{eq:eom}. Substituting the nonrelativistic approximations one may show that the spatial acceleration which follows from \eqref{eq:eom} is given by
\begin{equation}
\mathbf{a} = -\nabla \left( \Phi + \frac{\beta}{M_{\text{Pl}}} \delta\chi \right) \equiv -\nabla \Phi_{\text{eff}}.
\end{equation}
Using the relation $M_{\text{Pl}}^{-2} = 8\pi G$ we obtain the standard Yukawa-corrected potential
\begin{equation}
\label{eqphixi}
\Phi_{\text{eff}}(r) = -\frac{GM}{r} \left( 1 + \xi e^{-m_\chi r} \right),
\end{equation}
where we have defined the dimensionless strength parameter of the fifth force as $\xi \equiv 2\beta^2$. Equation \eqref{eqphixi} demonstrates that the effective gravitational potential experienced by a test body includes an attractive correction arising from the vacuum displacement generated by $M$. 

To assess the physical viability of this framework, we place initial constraints on its coupling strength using precision tests of the Weak Equivalence Principle (WEP). The Eötvös parameter $\eta_{\text{E}}$, which quantifies the fractional acceleration difference between test bodies of distinct compositions, is strictly bounded by $|\eta_{\text{E}}| < 1.1 \times 10^{-15}$ from the MICROSCOPE space mission\cite{MICROSCOPE:2022doy} and high-precision laboratory torsion balances\cite{Wagner:2012ui}.

To connect our model to these experimental bounds, we examine the behavior of the effective potential in the long-range regime. When the Compton wavelength of the scalar field is much larger than the characteristic experimental scale $r$ ($m_\chi^{-1} \gg r$), expanding the Yukawa exponential to leading order yields
\begin{equation}
e^{-m_\chi r} = 1 - m_\chi r + \mathcal{O}\left((m_\chi r)^2\right) \approx 1.
\end{equation}
Substituting this approximation into Eq.~\eqref{eqphixi}, the effective potential simplifies to a globally rescaled Newtonian potential:
\begin{equation}
\Phi_{\text{eff}}(r) \approx -\frac{GM}{r}(1 + \xi) \equiv -\frac{G_{\text{eff}}M}{r},
\end{equation}
where $G_{\text{eff}} = G(1 + \xi)$ is the dynamically renormalized gravitational constant.

As established in Section~\ref{sec:eom_wep}, the scalar interaction couples universally to bare rest mass ($T = -\rho$), preserving WEP for ideal point-like particles. For a macroscopic composite body, however, internal pressure, kinetic stresses, and field binding energies modify the stress-energy trace from its pure rest-mass value to $T= -\rho(1 - 3P/\rho)$. Modeling the interior of a test body as a continuum, we parameterize this departure by the average pressure-to-density fraction $f \equiv \langle 3P/\rho \rangle \sim \mathcal{O}(10^{-4}\text{--}10^{-3})$, which reflects the fractional contribution of nuclear and electromagnetic binding energy to the body's total mass\cite{Damour:1994zq}.

The scalar charge per unit mass for a body with trace fraction $f$ is given by
\begin{equation}
q \equiv \frac{\beta}{M_{\text{Pl}}} (1 - f).
\end{equation}
Now consider two macroscopic test bodies falling in the gravitational field generated by an external source mass $M$. From Eq.~\eqref{eq:eom}, their spatial accelerations are
\begin{equation}
\mathbf{a}_1 = -\nabla \Phi - q_1 \nabla \delta\chi, \quad \mathbf{a}_2 = -\nabla \Phi - q_2 \nabla \delta\chi.
\end{equation}
In the long-range limit ($m_\chi r \ll 1$), the gradient of the scalar field perturbation generated by the central mass evaluates to 
\begin{eqnarray}
\nabla \delta\chi \approx (\beta M / 4\pi M_{\text{Pl}} r^2)\hat{\mathbf{r}}.    
\end{eqnarray}
The differential acceleration between the two bodies thus reduces to
\begin{equation}
\Delta \mathbf{a} \equiv \mathbf{a}_1 - \mathbf{a}_2 = -(q_1 - q_2) \nabla \delta\chi = -\frac{\beta^2 M}{4\pi M_{\text{Pl}}^2 r^2} (f_2 - f_1) \hat{\mathbf{r}}.
\end{equation}
Using $M_{\text{Pl}}^{-2} = 8\pi G$ and defining the dimensionless parameter $\xi \equiv 2\beta^2$ -- which physically represents the ratio of the scalar fifth-force strength to the Newtonian gravitational force -- the differential acceleration simplifies cleanly to
\begin{equation}
\Delta \mathbf{a} = -\frac{\xi G M}{ r^2} (f_2 - f_1) \hat{\mathbf{r}}.
\end{equation}

The fractional difference in acceleration is measured experimentally via the Eötvös parameter $\eta_{\text{E}}$:
\begin{equation}
\label{etv}
\eta_{\text{E}} \equiv 2 \frac{|\mathbf{a}_1 - \mathbf{a}_2|}{|\mathbf{a}_1 + \mathbf{a}_2|} \approx {\xi} |f_1 - f_2|\equiv {\xi} \Delta f,
\end{equation}
where we have approximated the average acceleration as $|\mathbf{a}_1 + \mathbf{a}_2| \approx 2 GM / r^2$.
Applying the MICROSCOPE observational bound $|\eta_{\text{E}}| < 1.1 \times 10^{-15}$ \cite{MICROSCOPE:2022doy}, we obtain an upper bound on the unscreened fifth-force coupling:
\begin{equation}
\xi \lesssim \frac{ |\eta_{\text{E}}|}{\Delta f} \sim \frac{1.1 \times 10^{-15}}{10^{-3}} \implies \xi \lesssim 10^{-12}.
\end{equation}

This stringent constraint implies that unscreened vacuum displacement forces are strongly suppressed in local laboratory environments. However, non-linear scalar self-interactions in dense media may trigger screening mechanisms—such as Chameleon-type mass generation \cite{Khoury:2003aq, Brax:2004qh}—dynamically suppressing $\xi$ locally while permitting $\mathcal{O}(1)$ effects in diffuse astrophysical settings.

\section{Vacuum Solutions with Spherical Symmetry}
\label{sec:vacuum_solutions}

In the absence of matter sources ($T_{\mu\nu} = 0$), the gravitational field equations and the scalar field evolution equation must be evaluated for the vacuum substrate. We consider a static, spherically symmetric spacetime metric parameterized in standard diagonal form:
\begin{equation}
\label{eq:spherical_metric}
ds^2 = -f(r) dt^2 + \frac{1}{f(r)} dr^2 + r^2 \left( d\theta^2 + \sin^2\theta \, d\phi^2 \right).
\end{equation}
in which the scalar field profile $\chi(r)$ is connected to an energy-momentum tensor $V_{\mu\nu}(\chi)$. In the vacuum region ($T_{\mu\nu} = 0$), the modified Klein-Gordon equation \eqref{eq:modified_kg} simplifies to the sourceless static differential equation:
\begin{equation}
\label{eq:vac_kg}
\frac{1}{r^2} \frac{d}{dr} \left( r^2 f(r) \frac{d\chi}{dr} \right) - \frac{dU}{d\chi} = 0.
\end{equation}
For the Higgs-type potential $U(\chi) = \frac{1}{4}\lambda(\chi^2 - v^2)^2$, the potential extrema satisfy $U'(v) = 0$ with $U(v) = 0$ at the vacuum expectation value $\chi = \pm v$. Consequently, the unique regular solution to Eq.~\eqref{eq:vac_kg} obeying the asymptotic boundary condition $\chi(r) \to v$ as $r \to \infty$ is the uniform spatial configuration $\chi(r) = v$.
Substituting this result into the scalar energy-momentum tensor yields $V_{\mu\nu} = 0$ identically everywhere in the exterior space. The field equations thus collapse to the vacuum Einstein field equations $R_{\mu\nu} = 0$. By Birkhoff's Theorem, the unique spherically symmetric solution is the standard Schwarzschild metric:
\begin{equation}
f(r) = 1 - \frac{2GM}{r}.
\end{equation}

This derivation establishes that in the absolute absence of matter-vacuum interaction terms ($Q^\nu = 0$), the theory identically recovers the vacuum geometry of General Relativity. The displaced vacuum state ($\delta\chi \neq 0$) and its associated non-Newtonian dynamics emerge purely as a response to a non-zero source stress-energy trace $T$. Consequently, modified gravitational effects remain locally tethered to material mass distributions and cannot trigger spontaneous vacuum instabilities in empty space.

Although the background geometry remains Schwarzschild outside a central body, the motion of a test body is governed by the effective potential $\Phi_{\text{eff}}$ from Eq.~\eqref{eqphixi}. An orbiting body perceives a distance-dependent effective gravitational mass:
\begin{equation}
M_g(r) = M \left( 1 + \xi e^{-m_\chi r} \right),
\end{equation}
where $\xi \equiv 2\beta^2$. Here $M_g(r)$ represents the dynamical mass inferred from Keplerian orbital motion, which deviates from the central ADM mass $M$ due to the Yukawa displacement gradient. As $r \to \infty$, the Yukawa correction vanishes exponentially, restoring $M_g \to M$.

\subsection{Precession of Planetary Perihelia}
\label{subsec:precession}

The trajectory of a test body moving through the spherically symmetric vacuum displacement field exhibits an anomalous precession of its orbital perihelion. In standard General Relativity, the first-order relativistic perihelion shift per revolution is given by
\begin{equation}
\label{eq:delta_phi_gr}
\Delta \phi_{\text{GR}} = \frac{6\pi GM}{a(1-e^2)c^2},
\end{equation}
where $a$ denotes the semi-major axis and $e$ is the orbital eccentricity. In the present framework, the explicit scalar coupling $\beta$ introduces a secondary, distinct precession contribution mediated by spatial gradients of the scalar substrate. To extract the orbital dynamics from the theoretical framework developed in Sec.~\ref{sec:eom_wep}, we begin with the modified geodesic equation governing timelike worldlines:
\begin{equation}
\label{eq:modified_geodesic_peri}
\frac{d^2x^\alpha}{d\tau^2} + \Gamma^\alpha_{\mu\sigma} \frac{dx^\mu}{d\tau} \frac{dx^\sigma}{d\tau} = -\frac{\beta}{M_{\text{Pl}}} \left(\delta^{\alpha}_{~\nu} + u^\alpha u_\nu\right) \partial^\nu \chi.
\end{equation}
Restricting our analysis to static, spherically symmetric backgrounds in the equatorial plane ($\theta = \pi/2$), integration of Eq.~\eqref{eq:modified_geodesic_peri} yields two conserved quantities along the particle trajectory: the specific angular momentum $L$,
\begin{equation}
L = r^2 \frac{d\phi}{d\tau},
\end{equation}
and the modified specific energy $\mathcal{E}$ (per unit bare rest mass),
\begin{equation}
\mathcal{E} = (1+2\Phi) \left(\frac{dt}{d\tau}\right) \exp\left( \frac{\beta\chi}{M_{\text{Pl}}} \right).
\end{equation}
Contracting the four-velocity under the timelike metric normalization $g_{\mu\nu} u^\mu u^\nu = -1$, substituting $\frac{dt}{d\tau}$ and $\frac{d\phi}{d\tau}$ yields the exact first-order radial equation of motion:
\begin{equation}
\label{eq:radial_exact}
\left(\frac{dr}{d\tau}\right)^2 = \mathcal{E}^2 \exp\left( -\frac{2\beta\chi}{M_{\text{Pl}}} \right) - (1+2\Phi)\left(1 + \frac{L^2}{r^2}\right).
\end{equation}
To evaluate the orbital dynamics and compute the resulting perihelion shift, we now decompose the scalar field into its asymptotic vacuum expectation value and its localized perturbation, $\chi(r) = v + \delta\chi(r)$. Substituting this expansion into the exact radial equation \eqref{eq:radial_exact} and working to linear order in the perturbation $\delta\chi$ we obtain
\begin{eqnarray}
\label{eqrad}
\left(\frac{dr}{d\tau}\right)^2 \simeq \tilde{\mathcal{E}}^2 \left(1+\frac{\beta^2 M}{2M^2_{\text{Pl}}\pi r}e^{-m_\chi r} \right) - (1+2\Phi)\left(1 + \frac{L^2}{r^2}\right).    
\end{eqnarray}
where $\tilde{\mathcal{E}}={\mathcal{E}}e^{{-\beta v}/M_{Pl}}$.

To translate the radial motion into an orbital trajectory, we introduce the standard coordinate transformation $u(\phi) \equiv 1/r$ and use the specific angular momentum $L$ to rewrite the proper time derivative of the radius as $dr/d\tau=-L(du/d\phi)$.
Substituting these relations, along with the Newtonian potential $\Phi = -GMu$, equation (\ref{eqrad}) reduces to
\begin{equation}
\left(\frac{du}{d\phi}\right)^2 \simeq \frac{\tilde{\mathcal{E}}^2}{L^2} \left( 1 + \frac{\beta^2 M u}{2M_{\text{Pl}}^2 \pi} e^{-m_\chi/u} \right) - \frac{1 - 2GMu}{L^2} \left( 1 + L^2 u^2 \right).
\end{equation}

Differentiating the above equation with respect to $\phi$ we end up with the modified Binet equation governing the orbit:
\begin{equation}
\label{eq:binet_modified}
\frac{d^2u}{d\phi^2} + u = \frac{GM}{L^2} + 3GMu^2 + \frac{\tilde{\mathcal{E}}^2 \beta^2 M}{2 L^2 M_{\text{Pl}}^2 \pi} \frac{d}{du} \left( u \, e^{-m_\chi/u} \right).
\end{equation}
The second term on the RHS represents the standard general relativistic contribution while the final term encodes the Yukawa-suppressed vacuum displacement gradient. 

For a long-range scalar interaction where the Compton wavelength greatly exceeds the orbital scale ($m_\chi / u \ll 1$), the exponential factor approaches unity ($e^{-m_\chi/u} \approx 1$). In this case (\ref{eq:binet_modified}) reduces to 
\begin{eqnarray}
\label{eq:binet_longrange}
\frac{d^2u}{d\phi^2} + u = \frac{M}{L^2} \Big(G+ \frac{\tilde{\mathcal{E}}^2 \beta^2 }{2 M_{\text{Pl}}^2 \pi}\Big) + 3GMu^2 .  
\end{eqnarray}
Equation \eqref{eq:binet_longrange} reveals a crucial physical feature regarding the long-range limit of the vacuum displacement model. Because the vacuum correction term is independent of $u$ (and thus of $r$), it does not introduce the spatial variation required to drive a secular shift in the orbital ellipse. Instead, it acts purely as a constant shift to the right-hand side.
Consequently, a strictly long-range, unscreened scalar interaction does not generate anomalous planetary perihelion precession. This demonstrates that the coupling parameter $\xi$ cannot be constrained by planetary ephemeris data in this regime; rather, such bounds must rely on local equivalence principle tests -- such as those from the MICROSCOPE mission -- where differential accelerations isolate the underlying vacuum displacement forces.

\section{Steady-State Accretion of a Non-Relativistic Fluid}
\label{sec:steady_state_accretion}

To explore the astrophysical implications of the vacuum displacement framework in a dynamic environment, we further examine the steady-state, spherically symmetric accretion of a non-relativistic fluid onto a compact mass $M$. In standard General Relativity, this process is governed by Bondi-Michel accretion\cite{Bondi:1952ni,Michel:1972oeq}. In the present framework, the exchange of energy-momentum between the infalling fluid and the scalar vacuum substrate introduces two competing physical effects: an explicit fifth-force radial acceleration and a field-dependent modulation of the fluid's effective rest-mass density.

We consider a static background spacetime described by the exterior Schwarzschild metric, where the scalar field asymptotically settles to $\chi \to v$ as $r \to \infty$. In the non-relativistic regime far outside the event horizon ($r \gg 2GM$), the infalling fluid velocity satisfies $u \equiv -u^r \ll 1$, allowing us to set $f(r) \approx 1$.

Conservation of rest-mass is governed by the modified continuity equation (\ref{eq10}) derived from the longitudinal projection of the non-conservative interaction vector, 
\begin{eqnarray}
\nabla_\mu (\rho u^\mu) = \frac{\beta}{M_{\text{Pl}}} \rho u^\nu \partial_\nu \chi.    
\end{eqnarray}
For a steady-state, purely radial inflow, this relation reduces to:
\begin{equation}
\label{eq:acc_continuity}
\frac{1}{r^2} \frac{d}{dr} \left( r^2 \rho u \right) = \frac{\beta}{M_{\text{Pl}}} \rho u \frac{d\delta\chi}{dr},
\end{equation}
where $\chi(r) = v + \delta\chi(r)$. Integrating Eq.~\eqref{eq:acc_continuity} we define the exact conserved mass accretion rate across any spherical shell:
\begin{equation}
\label{eq:mod_mdot}
\dot{M} = 4\pi r^2 \left[ \rho(r) \exp\left( -\frac{\beta}{M_{\text{Pl}}} \delta\chi(r) \right) \right] u(r) = \text{constant}.
\end{equation}
Equation~\eqref{eq:mod_mdot} reveals that the physical fluid density is dressed by the vacuum displacement profile. 

Concurrently, the radial velocity profile is determined by the modified Euler equation which follows from (\ref{eq:eom}). In the pressureless dust limit ($p \ll \rho u^2$), the radial equation of motion simplifies to:
\begin{equation}
u \frac{du}{dr} = -\frac{GM}{r^2} - \frac{\beta}{M_{\text{Pl}}} \frac{d\delta\chi}{dr}.
\end{equation}
Using the linearized Yukawa potential profile $\delta\chi(r) = -\frac{\beta M}{4\pi r M_{\text{Pl}}} e^{-m_\chi r}$ (recalling $M_{\text{Pl}}^{-2} = 8\pi G$ and $\xi \equiv 2\beta^2$), the radial gradient evaluates to:
\begin{equation}
\frac{\beta}{M_{\text{Pl}}} \frac{d\delta\chi}{dr} = \frac{\xi GM}{r^2} \left( 1 + m_\chi r \right) e^{-m_\chi r}.
\end{equation}
The total radial acceleration thus becomes:
\begin{equation}
u \frac{du}{dr} = -\frac{GM}{r^2} \left[ 1 + \xi \left(1 + m_\chi r\right) e^{-m_\chi r} \right].
\end{equation}
Integrating from spatial infinity (where $u \to 0$ and $\delta\chi \to 0$) yields the modified free-fall velocity profile:
\begin{equation}
\label{eq:mod_vel}
u(r) \approx \sqrt{\frac{2GM(1+\xi)}{r}} \quad \text{for } m_\chi r \ll 1.
\end{equation}

Substituting the modified velocity profile \eqref{eq:mod_vel} back into the continuity equation \eqref{eq:mod_mdot}, we isolate the explicit radial profile for the fluid density:
\begin{equation}
\label{eq:acc_density_profile}
\rho(r) \simeq \frac{\dot{M}}{4\pi \sqrt{2GM(1+\xi)}} \, r^{-3/2} \exp\left( -\frac{\xi GM}{r} e^{-m_\chi r} \right).
\end{equation}

Equation~\eqref{eq:acc_density_profile} highlights a crucial theoretical interplay between the two distinct manifestations of Equivalence Principle violation within this framework: (i) the enhanced gravitational mass $M(1+\xi)$ accelerates the infalling fluid to higher velocities ($u \propto \sqrt{1+\xi}$), which acts via continuity to dilute the radial density profile by a factor of $(1+\xi)^{-1/2}$; and (ii) the position-dependent modulation of rest mass ($\delta\chi < 0$) introduces an exponential factor $\exp\left( -\frac{\xi GM}{r} e^{-m_\chi r} \right) < 1$, which systematically suppresses the inner accretion density relative to standard General Relativity.

From an observational perspective, because accretion luminosity and radiative cooling rates in systems such as X-ray binaries and active galactic nuclei scale quadratically with density ($\propto \rho^2(r)$) \cite{Rybicki:2004hfl}, this vacuum-induced density profile modulation provides a distinct astrophysical signature. Consequently, high-resolution X-ray observations of accreting compact objects offer a powerful, complementary pathway to constrain the vacuum displacement parameters $(\xi, m_\chi)$ independently of local laboratory tests and solar system ephemerides.

\section{Discussion and Conclusions}
\label{sec:discussion}

In this paper, we have presented the theoretical framework and local phenomenology of the Vacuum Displacement Principle, a modified gravitational model in which baryonic matter interacts directly with a dynamical vacuum substrate modeled by an ultra-light dark scalar field $\chi$ with a Higgs-type potential. By introducing a phenomenological covariant coupling proportional to the trace of the matter energy-momentum tensor, we formalize the physical concept that matter acts as a localized impurity, displacing the vacuum field away from its vacuum expectation value (VEV) $v$.
Projecting the non-conservative energy-momentum exchange along and orthogonal to a particle's four-velocity reveals two distinct dynamical manifestations. The transverse projection yields an attractive spatial fifth force, while the longitudinal projection generates a position-dependent modulation of a particle's inertial rest mass. While universal free fall is strictly preserved for fundamental point masses -- respecting the WEP -- the position dependence of the rest mass drives localized violations of the EEP. In the weak-field, non-relativistic limit, the coupled system yields an effective gravitational potential featuring a standard Yukawa-type correction. Testing this potential against local gravitational experiments establishes operational bounds on the coupling strength. Precision Eötvös parameter tests (such as the MICROSCOPE mission\cite{MICROSCOPE:2022doy}) strictly constrain the bare coupling to $\xi \lesssim 10^{-12}$ via composition-dependent differential accelerations. As demonstrated in our analysis of orbital mechanics, because a strictly long-range, unscreened scalar interaction acts as a constant offset in the Binet equation rather than a spatial modulation dependent on orbital radius, it does not drive a secular perihelion shift; consequently, planetary ephemeris data cannot constrain $\xi$ in this regime, cementing local equivalence principle tests as the primary observational benchmark. We have also demonstrated that in the absence of matter sources, the unique static, regular solution to the scalar field equation is the uniform spatial VEV configuration $\chi(r) = v$. Consequently, the scalar energy-momentum tensor vanishes identically, and the exterior geometry reduces strictly to the Schwarzschild metric of General Relativity by Birkhoff's theorem. This guarantees that vacuum displacement effects remain locally bound to mass sources and do not excite spontaneous vacuum instabilities in empty space. Finally, modeling steady-state, non-relativistic fluid accretion onto a compact object reveals a profound interplay between kinetic dilution and field-dependent mass variations. While the fifth force increases the infalling fluid velocity, the position-dependent reduction of rest mass introduces an exponential factor that systematically suppresses the density profile near the central source relative to standard General Relativity.

The phenomenology introduced in this paper offers future perspectives that we intend to examine in forthcoming works. In fact, the transition to galactic scales allows one to model the vacuum displacement profile in order to study galactic cores and stellar disks. By analyzing the resulting modified rotation curves, we aim to address the flat rotation curve problem without introducing cold dark matter particles, performing detailed fits against the SPARC database\cite{Lelli:2016zqa} to place stringent, observationally-driven bounds on the coupling parameters. Subsequently, we also intend to investigate the global, homogeneous, and isotropic cosmological dynamics of this framework. Using the FLRW metric, we will explore how the cosmic matter density drives the cosmological evolution of the background scalar field, providing a dynamical mechanism for late-time cosmic acceleration that acts as an effective, time-varying dark energy.

\section{Acknowledgments}

RM acknowledges financial support from
FAPERJ Grant No. E-$26/010.002481/2019$.

\end{document}